\documentclass[10pt, letterpaper]{article}
\usepackage[utf8]{inputenc}

\usepackage{amsthm,amsmath}
\usepackage[utf8]{inputenc} 

\usepackage{hyperref}

\hypersetup{
    colorlinks=true,
    linkcolor=blue,
    filecolor=magenta,      
    urlcolor=cyan,
    }

\usepackage{booktabs}
\usepackage{graphicx}
\usepackage[margin=2cm]{geometry}

\usepackage{algorithm}
\usepackage{algorithmic}

\usepackage{acronym}

\usepackage{ulem} 
\usepackage{subfigure}

\usepackage{soul}

\usepackage{xcolor}

\usepackage{anyfontsize}

\begin{document}

\title{Network psychometrics and cognitive network science open new ways for detecting, understanding and tackling the complexity of math anxiety: A review}

\author{
M. Stella$^{}$\\
\\
\footnotesize$^{}$CogNosco Lab, Department of Computer Science, University of Exeter, UK\\
\footnotesize $^{}$m.stella@exeter.ac.uk.}


\maketitle

\begin{abstract}

Math anxiety is a clinical pathology impairing cognitive processing in math-related contexts. Originally thought to affect only inexperienced, low-achieving students, recent investigations show how math anxiety is vastly diffused even among high-performing learners. This review of data-informed studies outlines math anxiety as a complex system that: (i) cripples well-being, self-confidence and information processing on both conscious and subconscious levels, (ii) can be transmitted by social interactions, like a pathogen, and worsened by distorted perceptions, (iii) affects roughly 20\% of students in 63 out of 64 worldwide educational systems but correlates weakly with academic performance, and (iv) poses a concrete threat to students' well-being, computational literacy and career prospects in science. These patterns underline the crucial need to go beyond performance for estimating math anxiety. Recent advances with network psychometrics and cognitive network science provide ideal frameworks for detecting, interpreting and intervening upon such clinical condition. Merging education research, psychology and data science, the approaches reviewed here reconstruct psychological constructs as complex systems, represented either as multivariate correlation models (e.g. graph exploratory analysis) or as cognitive networks of semantic/emotional associations (e.g. free association networks or forma mentis networks). Not only can these interconnected networks detect otherwise hidden levels of math anxiety but - more crucially - they can unveil the specific layout of interacting factors, e.g. key sources and targets, behind math anxiety in a given cohort. As discussed here, these network approaches open concrete ways for unveiling students' perceptions, emotions and mental well-being, and can enable future powerful data-informed interventions untangling math anxiety.

\end{abstract}

\section*{Introduction}

Anxiety is a distressing feeling aimed at avoiding a potential threat \cite{moore2015affect,ashcraft2007working}. By eliciting a sense of danger, anxiety contributes to avoiding negative experiences \cite{ashcraft2001relationships}. Imagine walking alone in a wild forest. The thought of meeting a predator would boost anxiety and favour a risk-averse behaviour, e.g. avoiding shady spots. In clinical populations, anxious feelings of distress can be triggered also in absence of a real threat \cite{mcelroy2018networks}. When the risk of encountering a wild predator is replaced by having to solve a mathematical equation, some people might still perceive high levels of anxiety \cite{ashcraft2002math}. A growing body of literature identifies \textit{math anxiety} as a pathological feeling of tension - different from general anxiety - which deeply inhibits access to memory and thus critically impairs thinking, learning and remembering mathematical knowledge \cite{ashcraft2007working,buckley2016understanding,khasawneh2021impact,moore2015affect,ashcraft2002math,wang2015math}. 

Math anxiety is shown to block a person’s reasoning when confronted with mathematical situations, even simple ones like performing basic calculations or understanding a simple plot \cite{buckley2016understanding}. This definition appeared back in the 1950s, with the pioneering work of Dreger and Aiken \cite{dreger1957identification}, who pointed out math anxiety as a key factor behind students’ attitudinal difficulty with maths. Subsequent works found out that math anxiety occurs at all educational levels, from primary school to Higher Education \cite{tytler2012student,khasawneh2021impact,ashcraft2007working,devine2018cognitive}. Overwhelmed by math anxiety, students often end up failing/repeating mathematics, statistics or physics courses relying on mathematical skills \cite{siew2019usinganxiety} and this translates into lower academic performance and more time required for completing a degree \cite{bja2019university,kargar2010relationship,wang2015math}, with annexed issues of increased economic burden and debt \cite{buckley2016understanding,espino2017mathematics}. Besides affecting students’ well-being and academic careers, math anxiety can solidify negative attitudes of closure towards mathematics \cite{tytler2012student} and thus discourage students from further pursuing careers in STEM \cite{ahmed2018developmental}. Since it is estimated that over 7 million new job positions will require math-related skills by 2025 \cite{oecd2018oecd}, these trends motivate an urgent need to better detect and overcome math anxiety \cite{ahmed2018developmental}. 

This review adopts a complex systems approach \cite{koopmans2016complex} to reviewing and understanding key sources, impact and detection techniques relative to math anxiety. Complexity stems from the fact that math anxiety does not affect a single typology of maths students and it is not limited to the emotional sphere \cite{tsiouplis2019rethinking,devine2018cognitive,khasawneh2021impact,luttenberger2018spotlight,mutodi2014exploring,ahmed2018developmental}. Crucially, math anxiety can be transmitted along social interactions \cite{maloney2015intergenerational} and it can affect cognitive processing at both conscious and unconscious levels in ways that we know relatively little about \cite{ashcraft2007working,khasawneh2021impact,shapiro2012role}. To better understand such a complex clinical pathology, affecting the mental well-being and career prospects of even well-performing students in multiple ways \cite{luttenberger2018spotlight,ahmed2018developmental}, quantitative models of knowledge are crucial \cite{tsiouplis2019rethinking,siew2020applications,stella2020italian,golino2021intelligence}. Merging behavioural and cognitive data together with models and quantitative techniques from multiple fields is key to complexity modelling \cite{koopmans2016complex,poquet2020forum} and can be a powerful weapon for understanding and reducing math anxiety, as discussed in this review. To this regard, particularly promising are novel techniques from cognitive data science \cite{siew2019cognitive,siew2019using,stella2019forma,golino2017exploratory,stella2020forma,borsboom2021psychonet}, which can map how individuals affected by math anxiety perceive and link together different aspects of their experiences in potentially distorted and stress-inducing ways.

\subsection*{Review outline}

This work is divided in two main parts. In the first part, a review of recent literature from educational psychology identifies sources and targets of math anxiety, outlining how even well performing students can be affected by such clinical condition. The review outlines how this pathology can: (i) stem from and get diffused along social interactions, and (ii) alter cognitive mechanisms of memory load management. The second part contains a review of psychometric scales, psychometric networks and cognitive network methods measuring math anxiety. The review of these quantitative tools has the potential to highlight key relationships and contexts both promoting and being promoted by math anxiety. Reviewing these techniques across fields is crucial: In the hands of professionals in Education, these psychological detection techniques can enable simple yet powerful data-informed interventions for detecting and fighting specific aspects of math anxiety.

\section*{Part 1: A socio-cognitive profiling of math anxiety}

Learning endeavours often include moderate levels of anxiety, due to a variety of elements, including the educational challenge of passing a test or the effort required for producing an essay \cite{spielberger2015nature,kubsch2021exploring}. This first part of the current work reviews the key elements that make math anxiety different from general anxiety. Attention is devoted to recent studies outlining math anxiety as a clinical pathology, affecting the cognition of even well-performing and adult students \cite{luttenberger2018spotlight}.

\subsection*{Math anxiety is an excessive response to math-related tasks}

When performing a learning task or an exam, moderate levels of anxiety are normal and can correlate positively with better academic performance \cite{ashcraft2001relationships,cassady2002cognitive,moore2015affect}. Moderate anxiety boosts the production of adrenaline in the brain and it activates a sequence of threat-reaction mechanisms that increase reactiveness and quick thinking \cite{moore2015affect}. These aspects help interpreting the original finding by Cassady and Johnson \cite{cassady2002cognitive} that moderate anxiety had a positive effect over test performance in a population of 168 undergraduate students. These results were replicated  by Wang and colleagues \cite{wang2015math}. However, the interplay between anxiety and performance was found to be non-linear. In fact, both works found that even stronger levels of anxiety created reactions of panic and ended up critically impairing, rather than enhancing, students’ performance \cite{cassady2002cognitive,wang2015math}, in agreement with other independent investigations \cite{ashcraft2001relationships,ashcraft2002math}. This negative effect has been replicated multiple times in students performing mathematical tasks \cite{ashcraft2007working,shapiro2012role,mutodi2014exploring}. 

Academic performance is not the only element influenced by math anxiety. A survey of 40 high-school students by Espino and colleagues \cite{espino2017mathematics} highlighted how math anxiety can drastically decrease students' confidence in their mathematical abilities. This lowered confidence translates into students actively avoiding math-related courses, with detrimental effects over the development of their mathematical literacy. Similar effects were found in University \cite{marshall2017addressing} and other high school settings \cite{hill2016maths,erturan2015investigation}. This indicates that widening the number of courses offering math-related skills might have only limited positive effects over promoting mathematical literacy \cite{foley2017math}. To boost computational thinking \cite{stella2020mapping,weintrop2016defining} and mathematical reasoning \cite{tytler2012student} a first crucial step would be detecting math anxiety in a given learning environment, including University settings \cite{bja2019university}.

\subsection*{Math anxiety affects also well-performing students}

As a debilitating negative emotional reaction towards maths, a preliminary hypothesis might be that math anxiety stems from poor academic performance \cite{dreger1957identification}. Recent studies show that this claim is not supported by the data. An investigation involving almost 1800 students in elementary and middle school settings \cite{devine2018cognitive} showed that 77\% of the students who experienced math anxiety turned out to be normal-to-high achievers on standard curriculum tests. Hill and colleagues \cite{hill2016maths} found that correlations between levels of math anxiety and school performance surfaced at later educational stages. In a cohort of 1104 students attending middle and high schools in Italy, the authors found weak negative correlations between arithmetic test scores and math anxiety. In a population of 90 adults, Douglas and LeFevre showed how basic number and spatial skills failed to predict the occurrence of math anxiety \cite{douglas2018exploring}, suggesting for the presence of more complex factors in addition to numeracy skills behind math anxiety. Foley and colleagues reviewed the link between math anxiety and academic performance through results from over 3,300 students in the 2012 edition of PISA tests (Program for International Student Assessment) \cite{foley2017math}. Although a weak negative correlation was found between math anxiety and academic performance, several countries in East Asia identified patterns where high-achieving students exhibited also higher levels of math anxiety (between 0.44 and 1.41 standard deviations above the mean \cite{foley2017math}). Analogous results were replicated by Yi and colleagues \cite{yi2020maths}. Foley and colleagues also underlined that countries like Japan or Switzerland exhibited close levels of average maths performance (536 and 531, respectively), yet Swiss students exhibited considerably lower levels of math anxiety compared to Japanese learners \cite{foley2017math}. These differences indicate how the prevalence of math anxiety in high performing students is a complex phenomenon, only weakly correlated and thus partially explained by academic performance. Additional evidence to this comes from the cognitive study of Beilock and colleagues \cite{beilock2005high}, who found that, in a population of 93 undergraduates, individuals with higher cognitive skills (i.e. working memory capacity) displayed stronger levels of math anxiety.

The above studies are relevant also for grasping a view about the prevalence of math anxiety across different educational systems. This clinical condition was detected in 63 out of 64 educational systems that participated in PISA 2012 and it affected roughly 17\% of the over 3,300 students involved in the studies \cite{foley2017math,yi2020maths}. In a country with advanced STEM programmes like the US, almost 93\% of adults reported having experienced math anxiety \cite{blazer2011strategies}. These quantitative results suggest that math anxiety is a global phenomenon that might be \textit{vastly undetected} in student populations when only academic performance is considered \cite{luttenberger2018spotlight}. 

To sum up, the above findings indicate that maths performance is only a partial, limited estimator for the prevalence of math anxiety in student populations. Whereas math anxiety can impact academic performance, the above results indicate that even well performing learners can experience high levels of distress \cite{foley2017math,beilock2005high,devine2018cognitive,douglas2018exploring,hill2016maths}. This underlines the need for detecting math anxiety via alternative approaches.

\subsection*{Math anxiety, social interactions and cognitive mechanisms of working memory}

As an emotional condition, math anxiety affects the mental well-being of an individual and influences their perception of themselves, the others and their environments \cite{ashcraft2001relationships,ashcraft2007working,khasawneh2021impact}. The process of understanding math anxiety should thus occur on multiple levels, adopting in parallel: (1) an ecological psychology approach \cite{martin2012switching,maloney2015intergenerational}, e.g. math anxiety can stem from social interactions, and (2) a cognitive framing \cite{eysenck2007anxiety}, e.g. math anxiety can stem from specific mechanisms of cognitive information processing. We will follow these two directions in the remainder of this section.

\subsubsection*{Math anxiety can be aggravated by and transmitted through social interactions}

Ecological psychology is a core branch of psychological studies indicating the relevance of social interactions in determining human personal, cognitive and emotional identities \cite{martin2012switching}. Several recent studies indicated how math anxiety is tightly connected with social interactions. Vast evidence from the literature report that social learning environments and educational curricula might be key sources of math anxiety in students \cite{martin2012switching,mutodi2014exploring,moore2015affect,khasawneh2021impact,marshall2017addressing}. Mutodi and Ngirande \cite{mutodi2014exploring} found that, in a random sample of 120 students, negative experiences of failure and classroom settings devoid of inclusiveness both boosted the prominence of math anxiety. The important role of teachers and teaching practice in favouring the growth of math anxiety was confirmed also by other recent studies \cite{moore2015affect,wilson2017maths}. Moore and colleagues \cite{moore2015affect} pointed out how the creation of positive learning environments - providing students with feasible goals and rewards - could make maths more attractive and less prone to induce anxiety. The authors underlined how instructors should focus on creating a learning experience that stimulates students to have a positive expectation about their learning, re-framing failure in maths as an enriching experience, useful for personal growth and improvement, rather than as a source of shame \cite{wilson2017maths}. In fact, Wilson showed how eliciting shame over poor academic performance bolstered math anxiety in classroom settings \cite{wilson2017maths}. Several works \cite{mutodi2014exploring,martin2012switching,khasawneh2021impact,spielberger2015nature,marshall2017addressing} underlined that a more positive perspective on maths failure should be built also in family settings, since parents were found to play an important part in mitigating or worsening children’s and teenagers’ fears about mathematics achievements \cite{yi2020maths,martin2012switching}. Maloney and colleagues \cite{maloney2015intergenerational} found evidence for \textit{a contagion} of math anxiety, which can be transmitted within a family from parents to children. Once infected with negative attitudes towards mathematics, children would grow up internalising stereotypical perceptions (e.g. "maths is hard" \cite{siew2019usinganxiety}) that could ultimately favour the prevalence of maths anxiety and maths disengagement, making it harder for teachers to promote math literacy \cite{tytler2012student,marshall2017addressing}. Maloney and colleagues also found that this negative inter-generational feedback loop was weakened whenever parents with high levels of math anxiety helped children less frequently \cite{maloney2015intergenerational}. Analogous negative effects of parents' attitudes over children's math anxiety were found also in other independent investigations (cf. \cite{martin2012switching,luttenberger2018spotlight}). Even simple pressure from parents and teachers was found to be a significant predictor of math anxiety in over 3355 students from 65 countries \cite{yi2020maths}. In particular, pressure stemmed from students perceiving that their educational role models, e.g. teachers and parents, valued performance more than other skills relative to self-confidence and conceptual understanding. In other words, performance anxiety \cite{marshall2017addressing,mcleod1989beliefs} was found to be promoted by distorted perceptions of role models, with exacerbating effects over math anxiety itself \cite{yi2020maths,mcleod1989beliefs}. Distorted perceptions of role models could also amplify negative attitudes towards a student's self-perceived confidence in math skills, which co-occurred together with enhanced levels of maths anxiety \cite{luttenberger2018spotlight}. 

The above findings underline the complexity of math anxiety as a global phenomenon, going well beyond educational places like classroom settings. Nonetheless, classroom interactions are found to contribute to math anxiety. Interactions between peers were found to transmit or worsen math anxiety by facilitating emotional states where students might feel inferior to their peers \cite{mutodi2014exploring,mcleod1989beliefs}. Social interactions between peers were also found to transmit and boost “stereotype threats” in maths \cite{shapiro2012role,johnson2012experimental}, in particular the myth of “male superiority” in achieving success in mathematics \cite{leyva2017unpacking}. In social psychology, stereotype threats are subconscious biases representing incomplete yet quick-to-grasp information about specific groups \cite{spencer2016stereotype,mcintyre2003alleviating,spencer1999stereotype}. These stereotypes can inform people’s behaviour on subliminal levels \cite{spencer1999stereotype}. People might be aware that some piece of knowledge about a group or category (e.g. girls) might be completely unfounded and thus be a myth/stereotype \cite{leyva2017unpacking} (e.g. girls are worse at maths than boys). And yet, by being aware of the existence of such stereotype, that group might concretely under-perform in tasks related to that sphere of knowledge \cite{spencer2016stereotype,johnson2012experimental,erturan2015investigation,mcintyre2003alleviating}. For instance, girls aware of their stereotypical portrayal as low-performing students in maths - compared to “mythical” better performances of white male students in maths \cite{leyva2017unpacking} - were found to perform worse than boys with analogous mathematical literacy in computational tasks \cite{shapiro2012role,johnson2012experimental}. This effect was attributed to anxiety stemming from subconscious processing of stereotypical expectations \cite{shapiro2012role} and was replicated in different contexts \cite{spencer1999stereotype,erturan2015investigation}. Crucially, a study over a sample of 144 students (75 women and 42 men) by Johns and colleagues \cite{johns2005knowing} found that raising awareness about stereotype threat could significantly improve students' computational performance, opening the way to future pedagogic interventions reducing such effect.

The above complex interplay between stereotype threat, self-perceived competence and differences in academic performance calls for the crucial need to highlight key cognitive mechanisms at work in learning environments, as potentially related to math anxiety and students' performance. 

\subsubsection*{A cognitive outlook on math anxiety}

In addition to the distorted perceptions of social agents and interactions that might aggravate math anxiety, one should also consider the key cognitive mechanisms at work in such clinical pathology \cite{ashcraft2002math,luttenberger2018spotlight}. Math anxiety alters cognitive performance in accessing, processing and producing information \cite{ashcraft2001relationships,ashcraft2007working}. All these aspects are strongly related to human memory, the cognitive system apt at storing and processing knowledge, stemming from neural activity in the brain circuitry \cite{vitevitch2019network,hillskenett2021}. From a cognitive perspective, human memory can be classified in different sub-components (cf. \cite{moore2015affect,ashcraft2001relationships,kumar2021semantic,montefinese2015looking}):
\begin{itemize}
    \item episodic memory deals with storing recollections of personal experiences;
    \item  semantic memory generates conceptual knowledge about the world and deals with the linguistic description of facts, ideas and events; 
    \item long-term memory deals with events that occurred in the distant past and are crucial for determining future behavioural patterns; 
    \item working memory stores and processes simple facts and bits of information that need to be combined while performing a computation.
\end{itemize}

Emotional processing \cite{montefinese2015looking}, personality traits \cite{beaty2021forward,lynn2020humans} and psycho-pathologies \cite{beilock2005high,castro2020contributions,castro2020quantifying} can heavily alter access and retrieval from these interconnected cognitive systems. Overwhelming evidence \cite{moore2015affect,buckley2016understanding,ashcraft2001relationships,ashcraft2007working,cassady2002cognitive} has shown that math anxiety in young adults (undergraduate students enrolled in Higher Education) can severely slow down recall of knowledge from semantic memory (e.g. remembering definitions or theorems) and also impair the maximum processing load from working memory (e.g. combining many entries in an equation in order to solve it).

Let us explore these two cognitive mechanisms through a computer science analogy. Like a computer with limited random access memory (RAM) for running applications, working memory in the human mind has a limited, finite capacity for processing information \cite{ashcraft2001relationships,cassady2002cognitive}. Such capacity is strongly affected by the attention level devoted to each task and by stimuli present in the environment \cite{beilock2005high,eysenck2007anxiety}. Attentional Control Theory \cite{eysenck2007anxiety} posits that anxiety can disrupt this balance between attention and environmental checks/stimulation. In general, working memory deals with processing not only information but also background tasks, e.g. sitting in a lecture room and paying attention to other colleagues or to the lecturer requires spatial coordination and reacting to others with emotional/memory responses \cite{ashcraft2002math}. Multiple tasks, including emotion regulation \cite{montefinese2015looking,pizzie2021association} or information processing \cite{cassady2002cognitive,beilock2005high}, have to be performed in parallel according to different capacities, as dictated by attention, up until maximum load is reached \cite{moore2015affect,ashcraft2007working,eysenck2007anxiety}. Through a review of several cognitive experiments, Moore et al. \cite{moore2015affect}, Buckley et al. \cite{buckley2016understanding} and Luttenberger et al. \cite{luttenberger2018spotlight} indicated that math anxiety can lower down the maximum capacity of working memory. This decrease corresponds to boosting resources allocated mainly to two processes: (i) emotion regulation \cite{gross2008emotion,pizzie2021association,montefinese2015looking}, i.e. a process apt at containing or managing switches between positive and negative emotional states,  and (ii) negative rumination, i.e. an abnormal persistence of negative thoughts affecting working, semantic and episodic memories \cite{gross2008emotion,buckley2016understanding}. In other words, current empirical studies and psychological coding theories of human memory converge towards a model in which students affected by math anxiety end up being overwhelmed with negative thoughts, which drain learners' mental capacity to perform computations \cite{ashcraft2001relationships,ashcraft2007working,cassady2002cognitive,beilock2005high,moore2015affect,ramirez2013math}. Notice that this emotion-related draining could be only weakly correlated with levels of domain knowledge of maths \cite{ashcraft2007working,ramirez2013math,kapuza2020network}, addressing why math anxiety has been reported to affect also well performing students (cf. \cite{devine2018cognitive} and \cite{ramirez2013math}). 

The above interplay underlines that math anxiety does not spawn from poor domain knowledge or competence only but it is rather a complex phenomenon that strongly depends on different emotional and information processing aspects. Hence, methods powered by cognitive psychology might considerably improve the detection, understanding and action against math anxiety.

\section*{Part 2: Innovative approaches in the literature for detecting math anxiety}

The previous Part outlined how math anxiety is a global, complex clinical condition affecting almost all educational systems across the globe and poorly predicted by academic performance \cite{khasawneh2021impact,martin2012switching,espino2017mathematics,devine2018cognitive,luttenberger2018spotlight}. These aspects underline how crucial it is to measure math anxiety via alternative mechanisms beyond performance itself \cite{yi2020maths,foley2017math}. Promoting measurements of math anxiety particularly in Higher Education would have mainly three beneficial effects: (i) it would improve students’ learning experience \cite{bja2019university}, (ii) freer from anxiety, students could be facilitated and encouraged to pursue successful careers in a growing field like STEM \cite{oecd2018oecd}, (iii) instructors could devise class-tailored techniques specifically targeting key aspects of anxiety in their classes. These points would disrupt a vicious bidirectional positive feedback loop, where increased math anxiety lowers mental capacity and academic performance, which in turn promote math anxiety itself \cite{foley2017math}. This disruption would contribute to the erasure of distorted, conflicting ideas about maths \cite{tytler2012student} and foster more creative connections between mathematics and other fields of knowledge discovery \cite{stella2019forma,stella2020mapping,weintrop2016defining}. Let us proceed with a review of novel psychological approaches to detecting and understanding maths anxiety mainly in terms of psychometrics \cite{golino2017exploratory,borsboom2021psychonet,haslbeck2021interpreting} and cognitive network science \cite{siew2019cognitive,hillskenett2021,stella2018multiplex}.

\subsection*{Psychometric approaches to detecting math anxiety}

Psychometrics is a key branch of psychology dealing with measuring psychological constructs like anxiety and, more in general, symptoms of psycho-pathologies through quantitative scales \cite{christensen2020psychometric,marsman2018introduction,golino2021investigating}. As a pathological state, math anxiety can be measured on a psychometric scale indicating the severity of the condition in a given individual. These scales are often self-assessed surveys \cite{hopko2003abbreviated,hunt2011development}, where students need to self-evaluate the overall severity of their anxiety and match their own experience against a given description, i.e. an item. For instance, an item might be “taking an examination in a math course” and students should have to rate how anxious they feel when in the situation described by such item, on a Likert scale from MIN (not anxious at all) to MAX (extremely anxious). Usually 5-point Likert scales are used, so that MIN equals 1 and MAX equals 5. The Abbreviated Math Anxiety Scale (AMAS) \cite{hopko2003abbreviated} outlined 20 different items relative to students’ learning of maths. The scale was validated through measures of internal consistency across samples. In a population of over 1239 undergraduate students enrolled in math-related programmes in Higher Education \cite{hopko2003abbreviated}, the items with stronger discriminative power for detecting high levels of math anxiety in self-reports were found to be “listening to a lecture in math class” and “taking an examination in a math course”. These items relate with the above connection between academic performance and math anxiety \cite{foley2017math}. 

Another scale commonly used for detecting math anxiety is the Mathematics Anxiety Scale-UK (MAS-UK) \cite{hunt2011development}. This scale included 23 items and was validated within a population of 283 British undergraduate students. Key groups of items (i.e. factors) explaining the most variance in the observed measures of math anxiety were: (i) self-evaluation of maths skills, (ii) calculations occurring in everyday life, and (iii) observation of math-related contexts and actors. These factors reflect the above discussed bidirectional influence that math anxiety has over reducing self-confidence in mathematical skills \cite{luttenberger2018spotlight,yi2020maths}.

\subsection*{Psychometric network science approaches to detecting math anxiety}

Items should not be considered as separate elements of students' experience but rather as deeply interconnected elements of potentially distorted perceptions \cite{borsboom2021psychonet,robinaugh2020network}. This is the key innovation of network psychometrics \cite{borsboom2021psychonet,robinaugh2020network,marsman2018introduction}, which models how individuals respond to specific items as a network combination of one or more latent variables, which cannot be measured directly by the experimenters. Hence, network psychometrics postulates the existence of a network structure driving the responses to items as a probability distribution associated with latent variables \cite{marsman2018introduction}. The selection of a model regulating responses to items is encoded by Item Response Theory \cite{reckase2009multidimensional}, which has fascinating analogues with the Ising model and the theory of interacting spins in statistical mechanics \cite{marsman2018introduction}. In physics, spins interacting on a given network topology have the tendency to align because of an underlying energy minimisation principle or in presence of an external magnetic field \cite{haslbeck2021interpreting}. When aligned, individual spins can create collective, global phenomena like non-zero magnetisation \cite{reckase2009multidimensional,haslbeck2021interpreting}. Analogously to spins aligned in a certain way, items might showcase specific responses (e.g. symptoms of anxiety). In presence of multiple items showcasing network-related responses, collective effects on the whole psychology of individuals might become evident (e.g. anxiety as a disorder) \cite{haslbeck2021interpreting}.

Identifying a specific network model relating item responses to collective clusters of symptoms is the key innovation behind \textit{graph exploratory analysis}, a technique introduced by Golino and Epskamp \cite{golino2017exploratory}. Graph exploratory analysis deals with estimating a limited number of dimensions that retain the highest explainable power of the variability observed in a large set of possibly correlated features, as expressed by items. Network estimation is performed through Markov Random Fields \cite{robinaugh2020network}, where nodes represent random variables connected by links indicating correlations or more general interaction metrics between node-level variables. Golino and Epskamp \cite{golino2017exploratory} adopt Gaussian graphical models \cite{robinaugh2020network} for capturing node interactions through an inverse co-variance matrix between normally distributed variables. In their model, network connections indicate the presence of a dependence between variables present after conditioning for all the other network variables. Once built, this network structure frames the problem of estimating psychological dimensions in terms of network community detection, i.e. finding clusters of tightly connected nodes \cite{tibshirani1996regression,dalege2017network,citraro2020identifying}. 

In more mathematical terms, Item Response Theory \cite{reckase2009multidimensional} assumes that the item responses $\textbf{y}_P $ produced by a person $P$  are the outcome of a linear transformation of $M$ latent variables, encoded in $\eta_P$, through a factor loading matrix $ \Lambda $ - to be estimated from the data - plus a random error term $\epsilon_P$:

\begin{equation}
    \centering
    \mathbf{y}_P= \Lambda \mathbf{\eta}_P + \mathbf{\epsilon}_P.
\end{equation}

Considering the variance-covariance matrix $\Psi = Var(\eta)$ and the diagonal matrix $\Theta = Var(\epsilon)$ (which assumes local independence), graph exploratory analysis \cite{golino2017exploratory} reconstructs network topology of correlations between items through the matrix $\textbf{K}$:

\begin{equation}
    \centering
    \textbf{K} = (\Lambda \Psi \Lambda^T + \Theta^{-1})^{-1},
\end{equation}
whose entries $k_{ij}$ indicate the local correlation between any two node-level variables $i$ and $j$. In fact, it can be shown analytically \cite{golino2017exploratory} that the off-diagonal elements of $\textbf{K}$ can be interpreted as partial correlation coefficients, whose clustering within communities can be identified through network science techniques like walktrap community detection \cite{golino2020eganet} or other recent approaches in community detection for feature-rich networks \cite{citraro2020identifying}.

The advantage of performing dimensionality/factor analysis through a complex network is in being able to immediately visualise and interpret key patterns of inter-dependencies through network science metrics \cite{borsboom2021psychonet}. This network approach to understanding psychological multivariate data is quickly growing within psychology \cite{borsboom2021psychonet,robinaugh2020network}, with pioneering applications in intelligence analytics \cite{golino2021intelligence}, estimation of key traits in attitudes from political discourse \cite{dalege2017network} and modelling of general anxiety symptoms in student \cite{siew2019usinganxiety} and non-student populations \cite{mcelroy2018networks}. The suitability of graph exploratory analysis in investigating data from students is supported by a recent study performed by Golino and colleagues \cite{golino2021investigating}. Through the social networks of 7 teachers, the authors investigated a population of 128 students between 2 and 11 years of age, reconstructing their concentration skills as an interconnected four-factor structure, i.e. emotion regulation, task engagement, empathy and imagination \cite{golino2021investigating}. An analogous adoption of graph exploratory analysis for investigating the key dimensions of math anxiety in specific student populations remains an unexplored gap, suitable for relevant future research. Interested readers might explore this methodology with an R package available online \cite{golino2020eganet} and capable of extracting factors from both multivariate psychometric data and text. 

A related approach is the one by Siew and colleagues \cite{siew2019usinganxiety}, who adopted network psychometrics for investigating stats anxiety in undegraduate students. Stats anxiety is an abnormal response to anticipating or performing statistical computations, analogous but not fully overlapping with math anxiety \cite{mallow2006science}. Siew and colleagues \cite{siew2019usinganxiety} built networks of items from the Statistical Anxiety Rating Scale and correlated them according to the responses of two subgroups of Higher Education students, suffering from high ($N_1$ = 115) and from low ($N_2$ = 113) stats anxiety. These two networks of item correlations displayed a difference of crucial relevance for math anxiety. Students suffering from higher stats anxiety organised their own responses to make the idea “maths is hard and necessary to excel in stats” considerably more central (in terms of network metrics like strength or closeness \cite{siew2019cognitive}) than what was found in students with lower stats anxiety. This finding was validated also in another study adopting a cognitive network approach \cite{siew2019using}. The evidence presented by Siew and colleagues \cite{siew2019usinganxiety} indicates that math anxiety might thus be a crucial aspect of stats anxiety, underlining the need to address and alleviate negative perceptions of mathematics across disciplines.

The above works underline how network psychometrics \cite{robinaugh2020network,golino2017exploratory} has recently produced a range of powerful techniques that could highlight and connect different complex aspects of math anxiety. This represents a fascinating, quantitative research direction that has been explored for other types of STEM anxiety \cite{siew2019usinganxiety} but not yet for math anxiety.

\subsection*{Cognitive network science approaches to detecting math anxiety}

Scales are powerful tools for investigating psychological constructs \cite{levy2021unveiling}. However they suffer from some limitations like potential biases and uncertainty in rating decisions \cite{calcagni2021psychometric}. Another key limitation of scales is that they rely on pre-determined items, that have to be identified \textit{a priori} by the experimenters \cite{hopko2003abbreviated}. However, the experience of specific groups of individuals might substantially differ and present aspects that might escape from a pre-determined/fixed list of items \cite{robinaugh2020network}. Differently put, a single list of items pre-determined by experimenters might not capture the full complexity of the ways specific elements are perceived, both cognitively and emotionally, by a given specific audience. 

This limitation can be overcome by accounting quantitatively for the different ways in which groups or individuals structure their own knowledge and emotions around concepts \cite{stella2020forma,correa2018word,stella2020lockdown}. Giving structure to knowledge is a paramount task of cognitive network science \cite{siew2019cognitive,castro2020contributions,lynn2020humans,hillskenett2021}, a research area dealing with modelling knowledge representations in the human mind as networks of interconnected conceptual elements, e.g. concepts linked by semantic \cite{kumar2021semantic}, syntactic \cite{guerreiro2021comparative} or phonological associations \cite{vitevitch2019network}. This field originated within psycholinguistics, where models of concept representations in the human mind employed network structures way before the advent of network science \cite{doczi2019overview}. Overwhelming theoretical and empirical research over the years has shown that mental representations of knowledge are highly structured \cite{doczi2019overview,vitevitch2019network} and such structure influences a variety of phenomena related to knowledge acquisition \cite{guerreiro2021comparative,stella2018distance} and processing \cite{lynn2020humans,lydon2021hunters,levy2021unveiling,castro2020quantifying}. For instance, the length of the shortest path between any two concepts in a network, i.e. network distance \cite{stella2018distance}, was shown to be predictive of normative language learning in young children \cite{stella2018distance,stella2018multiplex,vitevitch2019network}, creativity levels and curiosity in healthy populations \cite{stella2019viability,lydon2021hunters,beaty2021forward,kenett2018flexibility}, word production, picture naming and lexical access in semantic and working memory in healthy \cite{kumar2021semantic,doczi2019overview} or clinical populations \cite{castro2020quantifying,kenett2021creative,zaharchuk2021multilayer}. 

From the perspective of Education Science \cite{siew2020applications}, models of cognitive networks were used for mapping both the mindsets of educational agents (e.g. students \cite{siew2019using,stella2019forma,kapuza2020network}) and of educational supports (e.g. textbooks \cite{koponen2019lexical}). Koponen and Nousiainen \cite{koponen2019lexical} showed that two textbooks containing the same knowledge about physics, i.e. the same set of concepts, portrayed it with widely different connections, giving rise to different clusters of concepts and connections between them. Similarly, Kapuza and colleagues \cite{kapuza2020network} showed that the cognitive networks of expert and novice learners portrayed the same sets of concepts as connected in different ways, with experts identifying key concepts as being network nodes with more connections. Siew \cite{siew2019using} showed that networks of cue-target word recalls from semantic memory (i.e. free associations \cite{de2013better}) were found to be correlated with academic performance in a population of 104 undergraduate students \cite{siew2019using}. After controlling for network size, cognitive networks displaying larger network distances between concepts were associated with better academic performance \cite{siew2019using}. This indicated a positive trend between academic scores and being able to build long chains/cycles of conceptual associations in a mental representation of knowledge. Longer network distances in cognitive networks of memory recalls were found to characterise more semantically remote associations \cite{kumar2021semantic,valba2021analysis} and thus identified people with higher levels of creativity \cite{beaty2021forward,stella2019viability,kenett2018flexibility}. These approaches underline how much variety and insights can be encapsulated within conceptual associations as coming from different groups or sources. Cognitive network science provides a powerful framework for extracting information out of such associations \cite{siew2019cognitive,hillskenett2021,stella2020italian}.

The work of Stella and colleagues \cite{stella2019forma} implemented a quantitative detection of math anxiety via cognitive network science. The authors reconstructed associative knowledge and emotional perceptions around STEM concepts in two populations: (i) 159 high school students enrolled in math-heavy curricula and (ii) 59 STEM researchers. Memory recalls from semantic memory were extracted by participants through a cue-target continued free association game \cite{de2013better}. Participants were provided with a list of 50 words that were used as cues for stimulating semantic memory and they could respond to each cue by writing up to three free associates, e.g. reading “maths” made a participant immediately think of and react with “passion”, “creativity” and “hard”. These free associations, representative of semantic memory patterns \cite{kumar2021semantic,de2013better,kenett2018flexibility,kenett2021creative}, were enriched with emotional labels obtained from individual self-assessments (e.g. how positively would you rate the concept of “maths”?). Both these tasks required a simple spreadsheet and less than 1 hour for completion \cite{stella2019forma}. The resulting network linking cues and targets was coined as forma mentis (Latin for "mindset") network (FMN). Two FMNs were built, one representing students' and one representing STEM professionals' knowledge and emotions.

Figure 1 portrays the semantic/emotional frames of students (left) and STEM experts (right) around "mathematics". Students displayed an overwhelmingly negative perception of “maths” and associated it predominantly with other methodological ideas (e.g. “equation”, “formula”, “integral”, etc.). These concepts were rated by students as being mostly negative. Through cross-validation with an independent valence dataset by Fairfield and colleagues \cite{fairfield2017affective}, Stella and colleagues found that concentrations of negative associations corresponded to increased levels of anxiety. Not only maths, but also other STEM concepts like physics \cite{stella2019forma}, statistics \cite{stella2020italian} and computational thinking \cite{stella2020mapping} featured a concentration of negative associations and thus reported increased levels of anxiety. These trends were absent in the representations of knowledge coming from STEM experts. Hence, the quantitative reconstruction of students' mindsets via cognitive network science provided further evidence that math anxiety persists even in populations of students well trained in mathematics, in agreement with previous findings \cite{khasawneh2021impact,bja2019university,foley2017math,yi2020maths}. Anxiety did not affect students' perceptions of science \cite{stella2019forma}, teachers \cite{stella2020forma} or reasoning, in general \cite{stella2020mapping}. Negative relationships were almost completely replaced by positive ones in the representations of knowledge from STEM researchers, who rather associated maths with positive links to “creativity” and “art”. These cognitive associations were missing in the ways of thinking of students \cite{stella2019forma,stella2020forma}. Future pedagogic research could use such quantitative data for enhancing students' learning experience by favouring those specific creative conceptual associations present in the minds of STEM professionals \cite{stella2020mapping} but absent in student one's. Differently put, by reading the cognitive networks coming from STEM experts, education professionals might have an additional data compass for enhancing students' learning experiences and reducing math anxiety through specific associations.

\begin{figure}[t!]
\centering
\includegraphics[width=\textwidth]{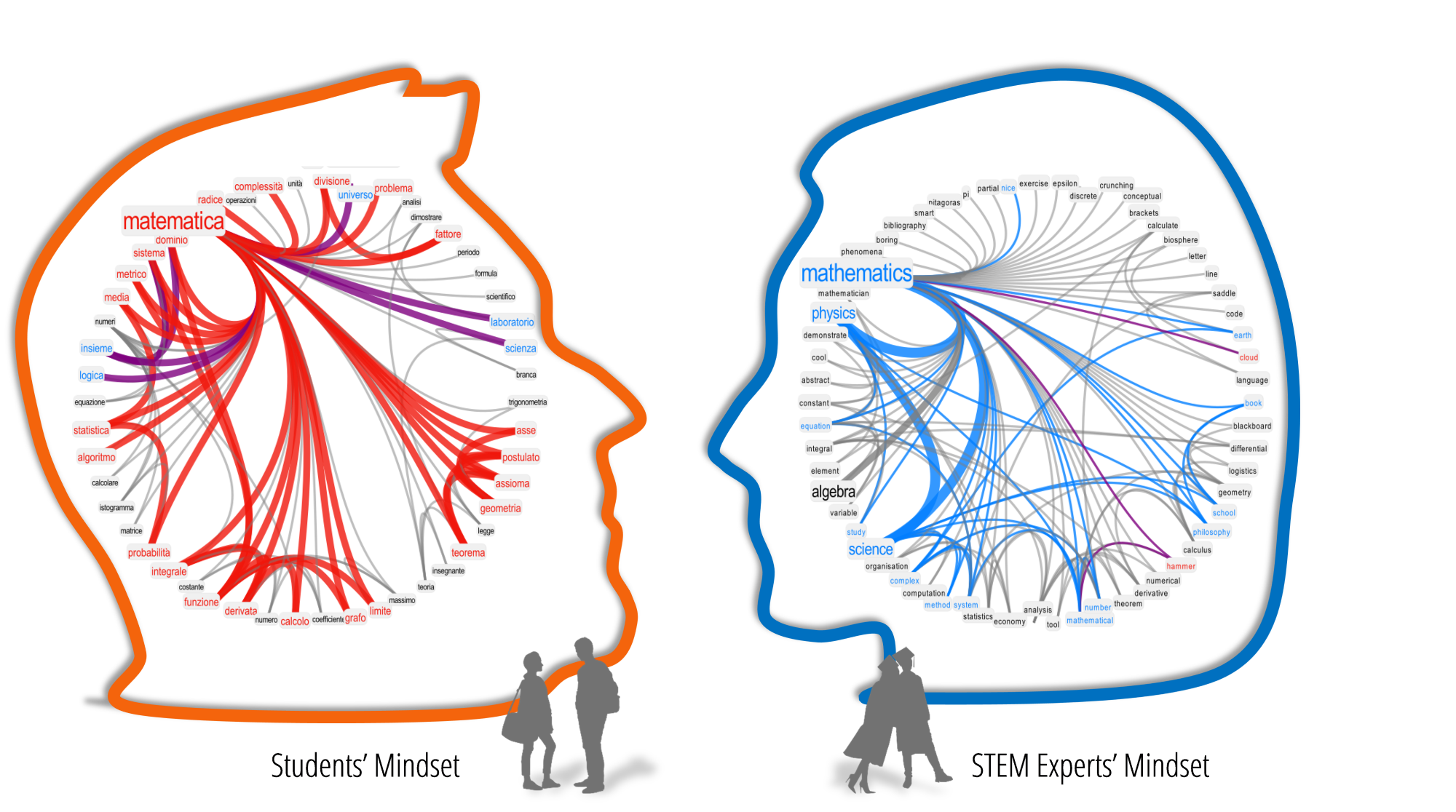}\label{Figure1.png}
\caption{Cognitive networks showing how students (left) and STEM experts (right) frame and perceive the concept of "mathematics". Colours indicate negative (red), positive (cyan) and neutral (black) valence perceptions as expressed by students and STEM experts. Links indicate memory recalls, showing which concepts are associated to "mathematics" in the minds of students and STEM experts.}
\label{fig:svo-main}
\end{figure}

\subsection*{Recommendations from the literature for fighting math anxiety}

To sum up, the above review indicates that math anxiety is a complex phenomenon, spreading along social interactions \cite{maloney2015intergenerational} and affecting memory, information processing and self-confidence of both low-achieving and well-performing students \cite{ashcraft2001relationships,ashcraft2007working,hill2016maths,yi2020maths}. Such complex and globally diffused phenomenon \cite{foley2017math} cannot be detected by considering academic performance only but it should rather be identified through psychological inquiries of students' minds \cite{siew2020applications}. Recent achievements from psychometrics \cite{golino2017exploratory,siew2019usinganxiety} and cognitive network science \cite{stella2019forma,stella2020forma,siew2019using} can be powerful tools up to the task. Psychometric scales can measure the gravity of math anxiety in students through a self-assessment of distressing experiences \cite{hopko2003abbreviated}, going well beyond simplistic detection of math anxiety based on performance \cite{yi2020maths}. Novel methods from psychometric network science like graph exploratory analysis \cite{golino2017exploratory,golino2020eganet} could shed light on the key dimensions of students' perceptions and experiences of math anxiety \cite{siew2019usinganxiety}. An analogous objective can be fulfilled by cognitive network science \cite{siew2019cognitive,castro2020contributions,hillskenett2021,stella2020forma}, which pointed out that math anxiety persists even in the mindsets of students enrolled in math-focused teaching curricula \cite{stella2019forma} and it is related to a dry perception of the discipline, lacking creative links present in the mindsets of STEM professionals \cite{stella2020italian}. Combined together, cognitive networks might identify evident and latent concepts in students' experience, which would then be identified as suitable items for building psychological constructs through graph exploratory analysis. The resulting cognitive multi-layer network structure \cite{de20218} could shed light on the organisation of key sources and targets of math anxiety in students, enabling a new generation of effective interventions. Notice that this synergy might include not only conceptual relationships, like in concept maps \cite{kapuza2020network}, but also feature affective/emotional perceptions and thus enable a complete conceptual/emotional profiling \cite{stella2020lockdown,fronzetti2020words} of students' mindsets. 

Both these techniques are simple enough that instructors might quickly use them for their own pedagogic practice. Forma mentis networks require simple spreadsheets and less than 1 hour for data gathering \cite{stella2019forma}. Cues might be selected as to replicate Stella and colleagues' studies \cite{stella2019forma,stella2020italian,stella2020forma,stella2020mapping} or to address specific needs of a given classroom. Graph exploratory analysis can be automated via a convenient R package \cite{golino2020eganet} and follow items provided by math anxiety scales \cite{hopko2003abbreviated,hunt2011development}. 

Using these models as a compass, teachers might then access not only what type of knowledge is available to their students but also what kind of positive or negative emotional states, like math anxiety or test anxiety \cite{erturan2015investigation,stella2020italian}, might be affecting their class in a given context. These cognitive/psychological networks might provide also quantitative landscapes of students' creativity and personalities \cite{valba2021analysis,kenett2018flexibility,beaty2021forward,hillskenett2021}, going well beyond student assessment based on quizzes only. Why would these analyses be useful? Letting students face their own emotions, as highlighted by network methodologies, and training them in emotional regulation would be a powerful way for reducing the negative effects of math anxiety, as suggested by recent studies \cite{pizzie2021association}.

In this way, a key question is relative to understanding more in detail which educational interventions might be guided by the above quantitative techniques. Let us discuss this aspect by concentrating over math anxiety and by following four distinct directions of relevance for students' learning \cite{koopmans2016complex}: (i) teaching environments, (ii) teaching styles, (iii) assessment methods and (iv) pedagogic innovation of maths curricula.

Teaching environments can affect the occurrence and gravity of math anxiety \cite{ashcraft2001relationships,ashcraft2007working,taylor2001mathematics,bja2019university}. Considering a blend of face-to-face and online teaching can have beneficial effects. By cutting the pressure of peer interactions and embedding students in environments as familiar as their very own houses, distance learning has been shown to drastically reduce math anxiety while favouring concentration over mathematical tools and reasoning \cite{taylor2001mathematics}. Analogous performance trends were found even during the current health crisis with COVID-19 \cite{spitzer2021academic}. Using psychometric/cognitive techniques to understand how students exactly perceive and frame distance learning, in relation to mathematics, could lead to innovative hybrid teaching environments, tailored around the needs of specific cohorts \cite{bja2019university,poquet2020forum}. Assessing the impact of such blended techniques, mixing face-to-face and remote learning, could open the way to substantially reducing math anxiety across different educational systems.

The psychometric results reviewed above converge towards the idea that math anxiety corresponds with a lowered self-confidence in maths skills \cite{siew2019usinganxiety,hopko2003abbreviated,hunt2011development}. Cognitive network science, instead, pointed out a dry perception of maths, where students think of mathematics as a purely dry technical discipline, devoid of creative or real-world applications \cite{stella2019forma,stella2020mapping}. Hence, teaching styles boosting self-confidence and creative aspects of maths literacy could have beneficial effects over reducing math anxiety. These styles could be designed and enhanced by education professionals sharing their "wisdom of practice" along networks of teaching practice \cite{mcleod1989beliefs,koopmans2016complex}. 

Assessments and tests can enhance anxiety, a phenomenon known as test anxiety \cite{mallow2006science,ashcraft2001relationships} and linked also to math anxiety \cite{hill2016maths}. Guided by the psychometric and cognitive analyses reviewed above \cite{siew2019using,hopko2003abbreviated,stella2020forma}, a way to reduce math anxiety could be devising more formative assessments, reducing the fear of failure and performance anxiety coming from peer pressure \cite{ashcraft2007working,khasawneh2021impact,kargar2010relationship}. More opportunities for students to measure their skills without performance anxiety should be considered. Flipped classroom or peer learning techniques from Education research \cite{marshall2017addressing,mattis2015flipped}, where students become active agents of their own learning, could be beneficial in implementing anxiety-free formative learning experiences \cite{bja2019university}.

Fighting math anxiety heavily relies on changing distorted perceptions of mathematics \cite{siew2019using,siew2019usinganxiety}. This could be tackled by promoting creative associations between specific concepts, as suggested by previous works \cite{stella2019forma,stella2020italian,ashcraft2002math}. Boosting students' learning progression through specific semantic frames or conceptual associations is proving successful in physics learning \cite{kubsch2021exploring} and it might prove a powerful technique also for enhancing the perception of maths. Hard mathematics is not always essential to master other subjects (see the above discussion of \cite{siew2019usinganxiety,stella2019forma,stella2020mapping}). The fact that specific groups might be better or worse at it is currently not supported by scientific evidence once subconscious biases are accounted for (cf. \cite{shapiro2012role}). Fighting these distorted perceptions can have great repercussions over eliminating the barrier posed by unconscious biases \cite{spencer1999stereotype,shapiro2012role,khasawneh2021impact}, ultimately favouring access to maths by a diverse set of students \cite{martin2012switching}. Channelling mathematics as a tool enabling a deeper understanding of our world can be a concrete way of disrupting maths stereotypes, as shown by the NetSciEd’s initiative in the United States \cite{cramer2015netsci,cramer2018network}. Academics and education professionals brought the mathematics of complex networks in high schools, involving both teachers and high schoolers into outreach programmes about understanding the real world with network science. Through a series of workshop events, scientific output and students’ feedback, analogous initiatives found that students translating mathematical concepts like matrices or equations into more concrete network representations ended up showing an enhanced appreciation of mathematics and decided to pursue a career in STEM (cf. \cite{cramer2018network}). The problem with underlining applications of a discipline like maths is that applied maths does not represent the complete landscape of mathematical knowledge. As cleverly summarised by Strogatz in \cite{strogatz2016writing}, mathematics contains an intrinsic beauty in its abstractness and universality, beyond its applications. This is why attention should be given towards presenting students with a more diverse landscape of maths achievements and relevance. Students should be presented with a complete profile of mathematical discoveries, both theoretical and applied, encompassing also historical background and relevance. By highlighting the struggles and achievements of mathematical physicists, Lommi and Koponen \cite{lommi2019network} found lectures about the history of science  solidified students' semantic memory and domain knowledge around curriculum subjects. Analogously, McIntyre and colleagues \cite{mcintyre2003alleviating} found that computational performance of women in academic tests could be enhanced by providing successful examples of women scientists, thus countering stereotype threat \cite{shapiro2012role} and providing contextual knowledge in classroom settings. These approaches provide evidence that merging mathematical knowledge with the history or context behind it might boost students' perceptions, reducing anxious and dry distorted mindsets \cite{stella2020forma,siew2020applications}.

\subsection*{Conclusions}

This multidisciplinary review outlined: (i) the socio-cognitive components of math anxiety, (ii) crucial key ways for measuring such pathological condition, and (iii) the most relevant investigations that provided evidence useful for fighting it. Reducing math anxiety in student populations is key to improving their mental well-being and preparing them better to the opportunities of the prospective job market. More research is required for understanding the unconscious/hidden sources of math anxiety, which is widely present even in well-performing learners \cite{devine2018cognitive,yi2020maths,foley2017math}. This makes academic performance a rather poor way for estimating the prevalence of math anxiety among students, a detection task that should and could be tackled by recent advancements from network psychometrics \cite{siew2019usinganxiety,golino2017exploratory}
and cognitive network science \cite{stella2019forma,siew2019using}. These methods are simple enough to be used in classroom settings and can crucially outline where math anxiety stems from and what kind of processes are drained by it. Access to such knowledge, tailored around specific cohorts and settings, can provide powerful guidance for effective strategies enhancing the appreciation of mathematics in the near generations of learners.

\bibliographystyle{plain}
\bibliography{sample}

\end{document}